\documentclass[12pt]{article}
\usepackage{amssymb}
\usepackage{amsmath}
\usepackage{graphicx,color}

\usepackage{graphics}
\usepackage{color}
\usepackage{amssymb,amsmath}
\usepackage{slashed}
\usepackage{epstopdf}
\usepackage{booktabs}
\usepackage{mathrsfs}
\usepackage[export]{adjustbox}
\usepackage{epsfig}
\usepackage{graphicx,color}

\newcommand{\bea}{\begin{eqnarray}}
\newcommand{\eea}{\end{eqnarray}}

\numberwithin{equation}{section}

\begin{document}
%
\begin{titlepage}
\begin{flushright}
YGHP17-02
\end{flushright}
\vspace*{10mm}
\begin{center}
\baselineskip 25pt 
{\Large\bf
Inflection-point inflation \\
in hyper-charge oriented
U(1)$_X$ model 
}

\end{center}
\vspace{5mm}
\begin{center}
{\large
Nobuchika Okada$^{~a}$, 
Satomi Okada$^{~b}$
and 
Digesh Raut$^{~a}$
}

\vspace{.5cm}

{\baselineskip 20pt \it
$^a$Department of Physics and Astronomy, University of Alabama, Tuscaloosa, AL35487, USA\\
$^b$Graduate School of Science and Engineering, Yamagata University,\\
Yamagata 990-8560, Japan\\
} 

\end{center}
\vspace{0.5cm}
\begin{abstract}
Inflection-point inflation is an interesting possibility to realize a successful slow-roll inflation
  when inflation is driven by a single scalar field with its value during inflation 
  below the Planck mass ($\phi_I \lesssim M_{Pl}$). 
In order for a renormalization group (RG) improved effective $\lambda  \phi^4$ potential
  to develop an inflection-point, the running quartic coupling $\lambda(\phi)$  must exhibit a minimum
  with an almost vanishing value in its RG evolution, namely $\lambda(\phi_I) \simeq 0$ 
  and $\beta_{\lambda}(\phi_I) \simeq 0$, where $\beta_{\lambda}$ is the beta-function of the quartic coupling. 
In this paper, we consider the inflection-point inflation in the context of 
  the minimal U(1)$_X$ extended Standard Model (SM),  
  a generalization of the minimal U(1)$_{B-L}$ model, where the U(1)$_X$ symmetry is realized as a linear combination of the SM U(1)$_Y$ and the U(1)$_{B-L}$ gauge symmetries. 
We identify the U(1)$_X$ Higgs field with the inflaton field. 
For a successful inflection-point inflation to be consistent with the current cosmological observations, 
   the mass ratios among the U(1)$_X$ gauge boson, the right-handed neutrinos 
   and the U(1)$_X$ Higgs boson are fixed. 
Focusing on the case that the U(1)$_X$ gauge symmetry is mostly oriented towards
   the SM U(1)$_Y$ direction, we investigate a consistency between the inflationary predictions 
   and the latest LHC Run-2 results on the search for a narrow resonance with the di-lepton final state.  
In addition, the inflection-point inflation provides a unique prediction for the running of the spectral index $\alpha \simeq  - 2.7 \times 10^{-3}\left(\frac{60}{N}\right)^2$ ($N$ is the e-folding number), which can be tested in the near future.

\end{abstract}
\end{titlepage}

\section{Introduction}

Inflationary universe is the standard paradigm in the modern cosmology \cite{inflation1, inflation2, chaotic_inflation, inflation3} 
   which provides not only solutions to various problems in the Standard Big Bang Cosmology, such as the flatness and horizon problems, 
   but also the primordial density fluctuations which seed the formation of large scale structure of the universe we see today. 
In a simple inflationary scenario known as slow-roll inflation, inflation is driven by a single scalar field (inflaton) 
   when it slowly rolls down to its potential minimum. 
During the slow-roll, the energy density of the universe is dominated by the inflaton potential energy, 
   which drives accelerated expansion of the universe, namely, cosmological inflation. 
After the end of inflation, the inflaton decays to Standard Model (SM) particles to reheat the universe 
   to initiate the Standard Big Bang Cosmology.

The slow-roll inflation requires the inflaton potential to be sufficiently flat in the inflationary epoch. 
In a simple inflationary scenario such as chaotic inflation, a flat potential is realized 
   by taking an initial inflaton value to be of the trans-Planckian scale.  
However, the field theoretical point of view, it may be more appealing to consider 
   the small-field inflation (SFI) scenario, where the initial inflaton value is smaller than the Planck mass 
   and possible higher-dimensional Planck suppressed operators are less important to the inflationary predictions. 
Hybrid inflation \cite{HybridInflation} is a well known example of the SFI \cite{HybridInflation},  
   where a flat direction of the scalar potential is realized with multiple scalar fields.       
When one considers the SFI driven by a single scalar field, 
   the so-called inflection-point inflation \cite{SF_Inf, InflectionPoint, InflectionPoint2} is an interesting possibility. 
If the inflaton potential exhibits an inflection-point, the slow-roll inflation epoch can be realized 
   with the initial inflaton value in the immediate vicinity of the inflection-point.

From a particle physics point of view,  an inflation scenario seems more compelling 
   if the inflaton field plays another important role in particle physics models,  
   such as the Higgs inflation scenario \cite{Higgs_inflation1,Higgs_inflation2,Higgs_inflation3} 
   in which the SM Higgs field is identified with the inflaton field. 
When the SM is extended with some extra gauge groups or unified gauged groups, 
   such models always include an extra Higgs field, in addition to the SM Higgs field,  
   to spontaneously break the gauge symmetries down to the SM one.   
Similarly to the Higgs inflation scenario, it is interesting if we can identify the extra Higgs field with the inflaton. 
The extra Higgs field usually has Yukawa couplings with some fermions in addition to the gauge
   and the quartic couplings, just like the SM Higgs doublet.  
As will be discussed below, this gauge-Higgs-Yukawa system is essential 
   to realize the inflection-point inflation with the identification of the Higgs field as the inflaton.

Let us consider a Renormalization-Group (RG) improved effective Higgs/inflaton potential~\cite{Sher}.  
During the inflation, we assume that inflaton value is much larger than its Vacuum Expectation Value (VEV) 
   at the potential minimum, so that the inflaton potential is dominated by its quartic term of the form,  
\bea
   V(\phi) = \frac{1}{4} \lambda(\phi) \;  \phi^4, 
\eea
   where $\phi$ denotes the inflaton field, and $\lambda(\phi)$ is the running quartic coupling.
If the RG running of the inflaton quartic coupling first decreases towards high energy and then increases, 
   inflection-point is realized in the vicinity of the minimum point of the running quartic coupling, 
   where both the quartic coupling and its beta-function become vanishingly small 
   \cite{InflectionPoint, InflectionPoint2}.\footnote{In the context of the $\lambda \phi^4$ inflation 
   with non-minimal gravitational coupling \cite{NonMinimalUpdate}, similar conditions 
   have been derived to ensure the stability of the inflaton potential \cite{RunningInflation1}.
}
In the vicinity of the inflection-point, the running quartic coupling obeys the (one-loop) RG equation of the form,  
\bea 
  16 \pi^2 \frac{d \lambda}{d \ln \phi} \simeq  C_g \; g^4  -  C_Y \; Y^4 ,
  \label{RGgeneral} 
\eea  
where $g$ and $Y$ are the gauge and Yukawa couplings, respectively, 
   and $C_g$ and $C_Y$ are positive coefficients whose actual values are calculable 
   once the particle content of the model is defined.  
Here, we have neglected terms proportional to $\lambda$ ($\lambda^2$ term and the anomalous dimension term) 
   because the SFI requires the quartic coupling $\lambda \propto g^6$, as will be shown later.  
Hence the quantum corrections to the effective Higgs potential are dominated by the gauge and Yukawa interactions.   
Realization of the inflection-point requires a vanishingly small beta-function 
   at the initial inflaton value, namely $C_g \;  g - C_Y \;  Y = 0 $.  
This condition leads to a relation between $g$ and $Y$, or in other words, 
    the mass ratio of gauge boson to the fermion in the Higgs model is fixed. 
Since the Higgs quartic coupling at low energy is evaluated by solving the RG equation, 
   the resultant Higgs mass also has a unique relation to the gauge and the fermion masses. 
Therefore, in the inflection-point inflation scenario with the Higgs field as the inflaton, 
   there is a correlation between the very high energy physics of inflation 
   and the low energy particle phenomenology.

Recently, two of the authors of this paper (N.O. and D.R.) have investigated the inflection-point inflation 
   in the minimal gauged $B-L$ (baryon number minus lepton number) extension of the SM \cite{mBL}, 
   where the $B-L$ Higgs field as the inflaton field~\cite{OR_SFI}. 
In order to realize the successful inflection-point inflation, we have obtained the predictions 
   for the mass spectrum for the $B-L$ gauge boson ($Z^\prime$ boson), 
   the right-handed neutrinos, and the $B-L$ Higgs boson 
   as a function of the initial inflaton value ($\phi_I$)  and the inflaton/Higgs VEV ($v_{BL}$). 
Considering the reheating after inflation with the fixed particle mass spectrum, 
   we have identified the allowed parameter regions to satisfy the Big Bang Nucleosynthesis constraint.  
We have found that  the entire parameter region for $m_{Z^\prime} \lesssim 500$ GeV can be tested 
   by the future collider experiments such as the High-Luminosity Large Hadron Collider (LHC) \cite{HL-LHC} 
   and the SHiP~\cite{SHIP} experiments.

In this paper, we generalize the minimal $B-L$ model to the so-called non-exotic 
   U(1)$_X$ extension of the SM \cite{Appelquist:2002mw}.  
The  non-exotic U(1)$_X$ model is the most general extension of the SM with an extra 
   anomaly-free U(1) gauge symmetry, which is described as a linear  
   combination of the SM U(1)$_Y$ and the U(1)$_{B-L}$ gauge groups. 
The particle content of the model is the same as the one in the minimal $B-L$ model 
   except for the generalization of the U(1)$_X$ charge assignment for particles.  
The orientation of the U(1)$_X$ gauge group is characterized by a U(1)$_X$ charge 
   of the SM Higgs doublet ($x_H$). 
For example, $x_H=0$ is the U(1)$_{B-L}$ limit, while the U(1)$_{B-L}$ gauge group 
   is oriented to the SM U(1)$_Y$ direction for $|x_H| \gg 1$.  
In this context, we investigate the inflection-point inflation with the identification 
   of the U(1)$_X$ Higgs field as the inflaton.   
As we will discuss in the following, the inflation analysis weakly depends on $x_H$  
   and hence our results are similar to those in Ref.~\cite{OR_SFI}.  
However,  there is a sharp contract in low energy phenomenologies, in particular, 
   the U(1)$_X$ gauge boson phenomenology at the LHC.  
An upper bound on the U(1)$_{B-L}$ gauge coupling, $g_{BL} \lesssim 0.01$, has been  
   obtained from theoretical consistencies in Ref.~\cite{OR_SFI}.  
The $Z^\prime$ boson with such a small coupling can be explored in future collider experiments 
    only for a small mass region such as $m_{Z^\prime} \lesssim 500$ GeV.   
For this case in the inflection-point inflation scenario, the reheating temperature 
    is estimated to be $T_R \lesssim 1$ GeV \cite{OR_SFI}. 
Such a low reheating temperature may not be desirable in terms of thermal dark matter physics  
    and baryogenesis scenario.   
On the other hand,  in the U(1)$_X$ generalization, the coupling of the $Z^\prime$ boson 
    with the SM fermions is controlled by $g_X x_H$ for $|x_H| \gtrsim 1$ 
    with $g_X$ being the U(1)$_X$ gauge coupling, and therefore the $Z^\prime$ boson coupling  
    becomes sizable for $|x_H| \gg1$. 
In this paper, we will investigate the inflection-point inflation for this hyper-charge oriented $U(1)_X$ extension of the SM, 
   which opens up a possibility to explore the mass region of $m_{Z^\prime} > 1$ TeV 
   at the LHC Run-2 while successfully realizing the  inflection-point inflation.

The paper is organized as follows. 
In the next section, we give a brief review of the slow-roll inflation. 
In Sec.~3, we present the inflationary predictions for the scenario, where the inflaton potential exhibits an inflection-point-like behavior during the slow-roll. 
In Sec.~4, we consider the minimally gauged $B-L$ extension of the SM, where the $B-L$ Higgs field is identified with the inflaton field. 
To realize the inflection-point in a Higgs/inflaton potential, we consider the RG improved effective Higgs/inflaton potential. 
In Sec.~5, we consider the constraints on the model parameters from the Big Bang Nucleosynthesis and the current collider experiments. 
We also discuss the prospects of testing the scenario in the future collider experiments,  such as the High-Luminosity LHC and SHiP experiments. 
Sec.~6 is devoted to conclusions.

\section{Basics of Inflection-Point Inflation}
The inflationary slow-roll parameters for the inflaton field ($\phi$) are given by 
\bea
\epsilon(\phi)=\frac{ M_{P}^2}{2} \left(\frac{V'}{V}\right)^2, \; \; 
\eta(\phi)=
M_{P}^2\left(\frac{V''}{V }\right), \;\;
\zeta^2{(\phi)} = M_{P}^4  \frac{V^{\prime}V^{\prime\prime\prime}}{V^2}, 
 \label{SRCond}
\eea
where $M_{P}= M_{Pl}/\sqrt{8 \pi} = 2.43\times 10^{18}$ GeV is the reduced Planck mass, 
   $V$ is the inflaton potential, and the prime denotes the derivative with respect to $\phi$.  
The amplitude of the curvature perturbation $\Delta^2_{\mathcal{R}}$ is given by 
\begin{equation} 
\Delta_{\mathcal{R}}^2 = \frac{1}{24 \pi^2}\frac{1}{M_P^4}\left. \frac{V}{ \epsilon } \right|_{k_0},
 \label{PSpec}
\end{equation}
  which should satisfy $\Delta_\mathcal{R}^2= 2.195 \times10^{-9}$
  from the Planck 2015 results \cite{Planck2015} with the pivot scale chosen at $k_0 = 0.002$ Mpc$^{-1}$.
The number of e-folds is defined as
\begin{eqnarray}
N=\frac{1}{M_{P}^2}\int_{\phi_E}^{\phi_I} d\phi \; \frac{V }{V^\prime}  ,
 \label{EFold}
\end{eqnarray} 
where $\phi_I$ is the inflaton value at a horizon exit corresponding to the scale $k_0$, 
  and $\phi_E$ is the inflaton value at the end of inflation, 
  which is defined by $\epsilon(\phi_E)=1$.
The value of $N$ depends logarithmically on the energy scale during inflation 
  as well as on the reheating temperature, and it is typically taken to be $50 - 60$.

The slow-roll approximation is valid as long as the conditions 
   $\epsilon \ll 1$, $|\eta| \ll 1$ , and $\zeta^2\ll1$ hold. 
In this case, the inflationary predictions are given by
\bea
n_s = 1-6\epsilon+2\eta, \; \; 
r = 16 \epsilon , \;\;
\alpha = 16 \epsilon \eta -24 \epsilon^2-2 \zeta^2, 
 \label{IPred}
\eea 
where $n_{s}$, $r$ and $\alpha \equiv \frac{\mathrm{d}n_s}{d ln k}$ are the scalar spectral index, 
  the tensor-to-scalar ratio and the running of the spectral index, respectively, which are evaluated at $\phi = \phi_I$.  
The Planck 2015 results \cite{Planck2015} set an upper bound on the tensor-to-scalar ratio as $r \lesssim 0.11$,  
   while the best fit value for the spectral index ($n_s$) and the running of spectral index ($\alpha$) are 
   $0.9655 \pm 0.0062$ and $ -0.0057\pm 0.0071$, respectively, at $68 \%$ CL.

In the SFI scenario, to realize the slow-roll inflation the inflaton potential must exhibit an inflection-point-like behavior, 
   where the potential is very flat.\footnote{
  For successful inflation scenario it is not necessary for the potential to realize an {\it exact} inflection-point. 
 We only require the inflaton potential to exhibit a behavior of almost an inflection-point.
}
Setting the inflaton value at the horizon in the very flat region $\phi_I = M$ of the potential, 
   we approximate the inflaton potential by the following expansion around $\phi = M$:\footnote{
Although our parameterization of the inflaton potential is slightly different, 
   most of analysis in this section overlaps with that in Ref.~\cite{InflectionPoint2}.       
}    
\bea
V(\phi)\simeq V_0 +V_1 (\phi-M)+\frac{V_2}{2} (\phi-M)^2+\frac{V_3}{6} (\phi-M)^3, 
\label{PExp}
\eea
where $V_0$ is a constant and $V_1$, $V_2$ and $V_3$ are the first, second and third derivatives 
  of the inflaton potential evaluated at $\phi=M$. 
When $V_1$ and $V_2$ are vanishingly small, the inflaton potential exhibits the inflection-point-like behavior. 
From Eqs.~(\ref{SRCond}) and (\ref{PExp}), the slow-roll parameters are then given by 
\bea
\epsilon(M) \simeq \frac{M_{P}^2}{2} \left( \frac{V_1}{V_0} \right)^2, \;\;
\eta(M) \simeq M_{P}^2 \left( \frac{V_2}{V_0} \right), \;\;
\zeta^2{(M)} = M_{P}^4  \frac{V_1 V_3}{V_0^2}, 
\label{IPa}
\eea
where we have used the approximation $V(M)\simeq V_0$. 
Similarly, the power-spectrum $\Delta_{\mathcal{R}}^2$ is expressed as
\bea
\Delta_{\mathcal{R}}^2 \simeq \frac{1}{12\pi^2} \frac{V_0^3}{M_P^6 V_1^2}.
\label{CV1} 
\eea 
Using the constraint $\Delta_{\mathcal{R}}^2= 2.195 \times  10^{-9}$ from the Planck 2015 results, 
   we can express $V_1$ and $V_2$ as 
\bea
\frac{V_1}{M^3}&\simeq& 1961\left(\frac{M}{M_P}\right)^3\left(\frac{V_0}{M^4}\right)^{3/2}, \nonumber \\
\frac{V_2}{M^2}&\simeq& -1.725\times 10^{-2} \left(\frac{1-n_s}{1-0.9655} \right) \left(\frac{M}{M_P} \right)^2 
   \left(\frac{V_0}{M^4}\right), 
\label{FEq-V1V2}
\eea
where we have used $V(M)\simeq V_0$ and $\epsilon(M) \ll |\eta(M)|$ as will be verified later. 
For the following analysis we set $n_{s}=0.9655$ at the center value from the Planck 2015 results \cite{Planck2015}.

We define the inflaton value ($\phi_E$) at the end of inflation by $\epsilon(\phi_E)=1$.  
Using Eq.~(\ref{EFold}), the e-folding number ($N$) is given by
\bea
N=\frac{2 V_0}{M_{P}^2\sqrt{-V_2^2 + 2 V_1 V_3} }\arctan \left(\frac{V_2 + V_3(\phi-M) }{\sqrt{-V_2^2+2 V_1 V_3}}\right) 
   \Big|_{\phi=M(1 - \delta_E)}^{\phi=M}, 
\label{CV3} 
\eea 
where we have parametrized $\phi_E$ as $\phi_E/M = 1- \delta_E $ with $0 < \delta_E <1$. 
The inflection-point-like behavior of the inflaton potential requires $V_1, V_2 \simeq 0$ and $V3  > 0 $,
  so that we can approximate $-V_2^2+2 V_1 V_3 \simeq 2 V_1 V_3$. 
This approximation is justified later. 
As we will also show later, $V_2,  \sqrt{2 V_1 V_3} \ll V3 M \delta_E $ and hence the e-folding number 
  is approximated as
\bea
N\simeq \frac{2 V_0}{M_{P}^2\sqrt{2 V_1 V_3}}  \arctan \left[\frac{V_3 M \delta_E }{\sqrt{2 V_1 V_3}}\right] 
  \simeq \pi \frac{V_0}{M_{P}^2\sqrt{2 V_1 V_3}}. 
\label{CV4} 
\eea
Using Eq.~(\ref{FEq-V1V2}), $V_3$ is then given by 
\bea
\frac{V_3}{M} \simeq 6.989 \times 10^{-7} \; \left( \frac{60}{N} \right)^2 
  \sqrt{\frac{V_0}{M^2 M_P^2}}. 
\label{FEq-V3} 
\eea 
From Eqs.~(\ref{FEq-V1V2}) and (\ref{FEq-V3}), 
  we find $2 V_1 V_3 \simeq 9.2 (60/N) V_2^2 $, and $-V_2^2+2 V_1 V_3 \simeq 2 V_1 V_3$  
  is a good approximation for $N=50-60$.

Using Eqs.~(\ref{IPred}), (\ref{IPa}), (\ref{FEq-V1V2}),  and (\ref{FEq-V3}), we now express 
   all inflationary predictions in terms $V_0$, $M$ and $N$. 
From Eqs.~(\ref{IPa}) and (\ref{FEq-V1V2}), the tensor-to-scalar ratio ($r$) is given by
\bea 
r=3.077\times 10^7 \; \left( \frac{V_0}{M_P^4} \right). 
\label{FEq-r}
\eea 
The running of the spectral index ($\alpha$) is found to be 
\bea
\alpha \simeq - 2\zeta^2(M) = -  2.742 \times 10^{-3}\left(\frac{60}{N}\right)^2.  
\label{FEq-alpha}
\eea
Note that this $\alpha$ value is a unique prediction of the inflection-point inflation 
   independently of $V_0$ and $M$.   
This prediction is consistent with the current experimental bound, $\alpha = - 0.0057\pm 0.0071$ \cite{Planck2015}. 
It is expected that the future experiments can reduce the error to $\pm0.002$ \cite{RunningSpectral}, 
  and therefore the prediction of the inflection-point inflation scenario can be tested in the future.

\section{The Inflection-point U(1)$_X$ Higgs Inflation}
\begin{table}[t]
\begin{center}
\begin{tabular}{c|ccc|c}
      &  SU(3)$_c$  & SU(2)$_L$ & U(1)$_Y$ & U(1)$_X$  \\ 
\hline
$q^{i}_{L}$ & {\bf 3 }    &  {\bf 2}         & $ 1/6$       & $(1/6) x_{H} + (1/3) x_{\Phi}$   \\
$u^{i}_{R}$ & {\bf 3 }    &  {\bf 1}         & $ 2/3$       & $(2/3) x_{H} + (1/3) x_{\Phi}$   \\
$d^{i}_{R}$ & {\bf 3 }    &  {\bf 1}         & $-1/3$       & $(-1/3) x_{H} + (1/3) x_{\Phi}$  \\
\hline
$\ell^{i}_{L}$ & {\bf 1 }    &  {\bf 2}         & $-1/2$       & $(-1/2) x_{H} - x_{\Phi}$    \\
$e^{i}_{R}$    & {\bf 1 }    &  {\bf 1}         & $-1$                   & $(-1)x_{H} - x_{\Phi}$   \\
\hline
$H$            & {\bf 1 }    &  {\bf 2}         & $- 1/2$       & $(-1/2) x_{H}$   \\  
\hline
$N^{i}_{R}$    & {\bf 1 }    &  {\bf 1}         &$0$                    & $- x_{\Phi}$     \\
$\Phi$            & {\bf 1 }       &  {\bf 1}       &$ 0$                  & $ + 2x_{\Phi}$  \\ 
\end{tabular}
\end{center}
\caption{
The particle content of the minimal U(1)$_X$ extended SM. 
In addition to the SM particle content ($i=1,2,3$), the three right-handed neutrinos  
  ($N_R^i$ ($i=1, 2, 3$)) and the U(1)$_X$ Higgs field ($\Phi$), which is identified with the inflaton, are introduced.   
The extra U(1)$_X$  gauge group is defined with a linear combination of the SM U(1)$_Y$   
  and the U(1)$_{B-L}$ gauge groups, and the U(1)$_X$ charges of fields are determined by 
  two real parameters, $x_H$ and $x_\Phi$.   
Without loss of generality, we fix $x_\Phi=1$ throughout this paper. 
}
\label{table1}
\end{table}

The model we will investigate is the anomaly-free U(1)$_X$ extension of the SM, 
  which is based on the gauge group SU(3)$_c \times$SU(2)$_L \times$U(1)$_Y \times$U(1)$_X$. 
The particle content of the model is  listed in Table~\ref{table1}.
The covariant derivative relevant to U(1)$_Y \times$ U(1)$_X$ is given by 
\begin{equation}
D_\mu \equiv \partial_\mu  
      - i ( g_1 Y + \tilde{g} Q_X ) B_\mu - i g_X  Q_X  Z^{\prime}_\mu, 
 \label{Eq:covariant_derivative}
\end{equation}
where $Y$ ($Q_X$) are U(1)$_Y$ (U(1)$_X$) charge of a particle, and the gauge coupling $\tilde{g}$ 
   is introduced in association with a kinetic mixing between the two U(1) gauge bosons. 
Although we set  $\tilde{g}$ zero at the $U(1)_X$ symmetry breaking scale, it is generated 
   through the RG evolutions.  
The particle content includes three generations of right-hand neutrinos $N_R^i$ 
  and a U(1)$_X$ Higgs field $\Phi$, in addition to the SM particle content.  
The U(1)$_X$ gauge group is identified with a linear combination of 
   the SM U(1)$_Y$  and the U(1)$_{B-L}$ gauge groups, 
   and hence the U(1)$_X$ charges of fields are determined by 
   two real parameters, $x_H$ and $x_\Phi$.   
Note that in the model the charge $x_\Phi$ always appears as a product 
   with the U(1)$_X$ gauge coupling and it is not an independent free parameter. 
Hence, we fix $x_\Phi=1$ throughout this paper. 
In this way, we reproduce the minimal $B-L$ model with the conventional charge assignment 
   as the limit of $x_H \to 0$.\footnote{
For $x_H=- 4/5$ and $x_H=-2$,
 the U(1)$_X$ gauge group can arise from breaking of the SO(10)
grand unified gauge group into the Standard Model one via the SU(5)$\times$U(1)$_X$ for the standard SU(5) and the flipped SU(5) subgroups of the SO(10), respectively. 
   }
The limit of $x_H \to +\infty~(-\infty)$ indicates that the U(1)$_X$ is (anti-)aligned to the SM U(1)$_Y$ direction.  
The anomaly structure of the model is the same as the minimal $B-L$ model 
  and the model is free from all the gauge and the gravitational anomalies 
  in the presence of the three right-handed neutirnos.

The Yukawa sector of the SM is extended to have 
\bea
\mathcal{L}_{Yukawa} \supset  - \sum_{i=1}^{3} \sum_{j=1}^{3} Y^{ij}_{D} \overline{\ell^i_{L}} H N_R^j 
          -\frac{1}{2} \sum_{k=1}^{3} Y_M^k \Phi \overline{N_R^{k~C}} N_R^k 
       + {\rm h.c.} ,
\label{Lag1} 
\eea
where the first and second terms are the neutrino Dirac Yukawa coupling and the Majorana Yukawa coupling, respectively. 
Without loss of generality, the Majorana Yukawa couplings are already diagonalized in our basis.  
Once the U(1)$_X$ Higgs field $\Phi$ develops non-zero VEV,  
   the U(1)$_X$ gauge symmetry is broken and the Majorana masses for the right-handed neutrinos are generated. 
Then, the seesaw mechanism~\cite{Seesaw} is automatically implemented in the model 
  after the electroweak symmetry breaking.

The renormalizable scalar potential for the SM Higgs doublet ($H$) and the U(1)$_X$ Higgs ($\Phi$) fields is given by 
\bea  
V = \lambda_H \left(  H^{\dagger}H -\frac{v_h^2}{2} \right)^2
+ \lambda_{\Phi} \left(  \Phi^{\dagger} \Phi -\frac{v_X^2}{2}  \right)^2
+ \lambda_{\rm mix} 
\left(  H^{\dagger}H -\frac{v_h^2}{2} \right) 
\left(  \Phi^{\dagger} \Phi -\frac{v_X^2}{2}  \right) , 
\label{Higgs_Potential}
\eea
where all quartic couplings are chosen to be positive. 
At the potential minimum, the Higgs fields develop their VEVs as 
\bea
  \langle H \rangle =  \left(  \begin{array}{c}  
    \frac{v_h}{\sqrt{2}} \\
    0 \end{array}
\right),  \;  \;  \; \; 
\langle \Phi \rangle =  \frac{v_X}{\sqrt{2}}. 
\eea
Associated with the U(1)$_X$ symmetry breaking  (as well as the electroweak symmetry breaking),  
  the U(1)$_X$ gauge boson ($Z^\prime$ boson), the right-handed Majorana neutrinos, 
  and the U(1)$_X$ Higgs boson ($\phi$)  acquire their masses as 
\begin{eqnarray}
  m_{Z^\prime} = g_X \sqrt{4 v_X^2+  \frac{x_H^2}{4} v_h^2} \simeq 2 g_X v_X , 
   \;  \;  m_{N^i} = \frac{1}{\sqrt{2}} Y_M^i v_X, 
   \; \;  m_\phi = \sqrt{2 \lambda_\Phi} v_X, 
\end{eqnarray} 
where $v_h=246$ GeV is the SM Higgs VEV, and we have used the LEP constraint \cite{LEP} 
  $v_X^2 \gg v_h^2$. 
Because of the LEP constraint, the mass mixing of the $Z^\prime$ boson with the SM $Z$ boson
   is very small, and we neglect it in our analysis in this paper.

We identify the physical U(1)$_X$ Higgs field $(\phi)$ with the inflaton, 
  which is defined as $\Phi = (\phi+v_h)/\sqrt{2}$ in the unitary gauge,  
  and consider the inflation trajectory $\phi \gg v_X, v_h$ and $H=0$ 
    in the scalar potential in Eq.~(\ref{Higgs_Potential}). 
In our analysis, we employ the RG improved effective potential along this inflation trajectory, 
   which is dominated by the inflaton quartic term and given by
\bea
V(\phi) = \frac{1}{4} \; \lambda (\phi) \; \phi^4, 
\label{VEff}
\eea
where $\lambda (\phi)$ is the solution to the RG equation listed in Appendix.  
With the RG improved effective potential, we express the coefficients in the expansion of Eq.~(\ref{PExp}) as 
\bea
\frac{V_1}{M^3}&=& \frac{1}{4} \left(4 \lambda_\Phi + \beta_{\lambda_\Phi} \right),\nonumber \\
\frac{V_2}{M^2}&=& \frac{1}{4} \left(12\lambda _\Phi+ 7\beta_{\lambda_\Phi}+M \beta_{\lambda_\Phi}^\prime \right), \nonumber \\
\frac{V_3}{M}&=& \frac{1}{4}  \left(24\lambda_\Phi + 26\beta_{\lambda_\Phi}+10M \beta_{\lambda_\Phi}^\prime
  +M^2 \beta_{\lambda_\Phi}^{\prime\prime} \right), 
\label{ICons2}
\eea
where the prime denotes $d/d\phi$, and $\beta_{\lambda_\Phi}$ is 
   the beta function of the quartic coupling $\lambda_\Phi$ given by 
\bea
\beta_{\lambda_\Phi} &=&\frac{1}{(4\pi)^2} \left[ 
 \lambda_\Phi \left\{ 20\lambda_\Phi + 2\sum_{i=1}^3 (Y_M^i)^2 - 48 \left( \tilde{g}^2 + g_X^2 \right) \right\}
  +2\lambda_{\rm mix}^2  \right. \nonumber \\ 
&+& \left.  96\left( \tilde{g}^2 + g_X^2 \right)^2 - \sum_{i=1}^3 (Y_M^i)^4 
	\right],  
\label{RGEs}
\eea   
Using $V_1/M^3\simeq 0$ and $V_2/M^2\simeq 0$ to realize the inflection-point like behavior 
  of the inflaton effective potential, we obtain
\bea
 \beta_{\lambda_\Phi} (M)\simeq -4\lambda_\Phi(M), \qquad
 M \beta_{\lambda_\Phi}^{\prime}(M)\simeq 16 \lambda_\Phi (M). 
 \label{Cond1}
\eea
For the couplings being in the perturbative regime, we evaluate  
    $M^2 \beta_{\lambda_\Phi}^{\prime\prime}(M) \simeq - M \beta_{\lambda_\Phi}^{\prime}(M) \simeq -16 \lambda_\Phi(M)$ 
    as a good approximation. 
Hence the last equation in Eq.~(\ref{ICons2}) is simplified to be $V_3/M \simeq 16 \;\lambda_\Phi(M)$  
   and  Eq.~(\ref{FEq-V3}) leads to 
\bea
\lambda_\Phi(M)\simeq 4.770 \times 10^{-16} \left(\frac{M}{M_{P}}\right)^2\left(\frac{60}{N}\right)^4,
\label{FEq1} 
\eea 
where we have used $V_0\simeq (1/4) \lambda_\Phi(M) M^4$. 
For $M \lesssim M_P$, $\lambda_\Phi(M)$ is found to be very small.

In evaluating $\beta_{\lambda_\Phi}(M)$, we simply assume $\lambda_{\rm mix}(M)$ is negligibly small.  
Although we have set $\tilde{g}(v_X)=0$, non-vanishing $\tilde{g}(M)$ is generated through its RG evolution 
   through its beta function consisting of two terms: one is proportional to $\tilde{g}$ 
   and the other is proportional to $g_X$. 
However, as we will see later, the inflection-point inflation requires $g_X(M) \ll1$, 
  and hence $\tilde{g}$ stays negligibly small at any scales. 
As a result, Eq.~(\ref{Cond1}) with Eq.~(\ref{FEq1}) leads to $\beta_{\lambda_\Phi}(M) \simeq 0$, and we find 
\bea 
   Y_M (M)  \simeq 32^{1/4} g_X(M),  
\label{FEq2}   
\eea
where we have taken the degenerate Yukawa couplings for three right-handed neutrinos 
   $Y_M \equiv Y_M^1=Y_M^2=Y_M^3$, for simplicity. 
Therefore, the mass ratio between the right-handed neutrinos and the $Z^\prime$ gauge boson 
   is fixed to realize a successful inflection-point inflation.

Now RG equations for $\lambda_\Phi$, $g_X$ and $Y_M$ at the 1-loop level 
  are approximately give by 
\bea
16 \pi^2  \frac{d \lambda_\Phi}{d \ln \phi} &\simeq&  
     96 g_X^4   - 3 Y_M^4,  \nonumber \\
 16 \pi^2 \frac{d g_X}{d \ln \phi } &=&  
   \left( \frac{72 + 64 x_H + 41 x_H^2}{6} \right) g_X^3, 
  \nonumber \\
  16 \pi^2  \frac{d Y_M}{d \ln \phi} & \simeq & 
	Y_M \left(  \frac{5}{2} Y_M^2 -6  g_X^2 \right) .
\label{ApproxRGE}  
\eea
Using the second equation in Eq.~(\ref{Cond1}) and Eqs.~(\ref{FEq1})-(\ref{ApproxRGE}), 
   we express the U(1)$_X$ gauge coupling at $\phi=M$ as 
\bea
g_X(M)\simeq  \frac{ 1.511\times 10^{-2}}{\left(93+256x_H+164 {x_H}^2\right)^{1/6}}  
 \;\left(\frac{M}{M_{P}}\right)^{1/3}\left(\frac{60}{N}\right)^{2/3}.
\label{FEq3} 
\eea
Finally, from Eqs.~(\ref{FEq-r}) and (\ref{FEq1}), the tensor-to-scalar ratio ($r$) is given by 
\bea
r \simeq 3.670 \times 10^{-9}  \Big(\frac{M}{M_{P}}\Big)^6, 
\label{FEqR} 
\eea
which is very small, as expected for the SFI scenario.

At the end of inflation, $\epsilon(\phi_E)$ is explicitly given by
\bea
\epsilon(\phi_E) = \frac{{M_P}^2}{2 V_0^2} \left(V_1-V_2 \;M \delta_E+\frac{V_3}{2}M^2 
{\delta_E}^2\right)^2 \simeq   \frac{{M_P}^2\;  M^6\;  {\delta_E}^2}{2\; V_0^2 } 
\left(-\frac{V_2}{M^2} +\frac{V_3}{2 M} {\delta_E}\right)^2. 
\label{epsilon}
\eea  
We evaluate $\delta_E$ from $\epsilon(\phi_E)=1$. 
If we assume that the first term dominates in the parenthesis of the final expression above, we find  $\delta_E\gg 1$ by using Eqs.~(\ref{FEq-V1V2}), (\ref{FEq-V3}), and (\ref{FEq1}), which is inconsistent with $0 < \delta_E < 1$. 
Therefore, the second term must dominate, and hence we obtain
\bea
\delta_{E} \simeq 0.210 \; \Big(\frac{M}{M_{P}}\Big)^{1/2}, 
\label{delta}
\eea 
by using Eqs.~(\ref{FEq-V3}) and (\ref{FEq1}).

Before presenting our numerical results, let us check the consistency of our analysis. 
In the previous section we have approximated the inflaton potential by Eq.~(\ref{PExp}), neglecting the higher order terms. 
For consistency, we need to check if the contribution of higher order terms can actually be neglected in our model. 
Consider the following expansion of inflaton potential at $\phi=M$, 
\bea
V(\phi) = \sum_{n=0} \frac{V^{(n)}}{n !} (\phi-M)^n, 
\label{PExpGen} 
\eea
where $V^{(n)}$ is the $n$-th derivative of the potential evaluated at $\phi = M$. 
As has been discussed in the previous section, 
    $V_1=V^{(1)}$ and $V_2=V^{(2)}$ are fixed by the  experimental values 
     of the scalar power-spectrum ($\Delta_\mathcal{R}^2$) and the spectral index ($n_s$), respectively. 
For the consistency of our previous analysis, we require that the terms $V^{(4)}$ and higher 
     contribute negligibly in determination of $\delta_E$ compared to $V_3$ at the end of inflation. 
Using Eqs.~(\ref{SRCond}) and (\ref{PExpGen}), $\epsilon(\phi_E)$ is expressed as
\bea
\epsilon(\phi_E) &\simeq& \frac{{M_P}^2}{2 V_0^2} \left(\sum_{n=1} \frac{V^{(n)}}{(n-1)!} (\phi-M)^{n-1}\right)^2 \\ \nonumber
&\simeq& 
\frac{{M_P}^2}{2 V_0^2} \left( \frac{V_3}{2}M^2 {\delta_E}^2+ \sum_{n=4} \frac{V^{(n)}}{(n-1)!} (M \; \delta_E )^{n-1}\right)^2,  
\label{DelepsBound} 
\eea 
where we have used $V(\phi_E) \simeq V_0$. 
This leads to constraint 
\bea
{\delta_E}^{(p-3)} <  \left|\frac{(p-1)!}{2} \frac{V^{(3)}}{V^{(p)}} M^{3-p}\right|. 
\label{DelBound} 
\eea
where $p\geq4$. 
To proceed further we need to evaluate Eq.~(\ref{DelBound}) explicitly for the minimal U(1)$_X$ model.
As has been shown previously in this section, all the higher order derivatives of the potential can be approximately given by $V^{(n)} M^{n-4} \simeq C_n \lambda(M)$, where $C_n$ is a constant. 
For example, $C_4 = 96$ and $C_5=184$.  
We find that the most severe bound for both cases is from $V^{(4)}$ term.  
Using Eqs.~(\ref{delta}) we obtain an upper bound on $M< 5.67 M_{P}$.

\begin{figure}[h]
\begin{center}
\includegraphics[scale =1.6]{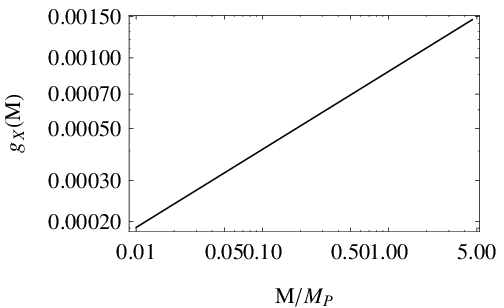} \;\;\;\;\;\;
\includegraphics[scale=1.55]{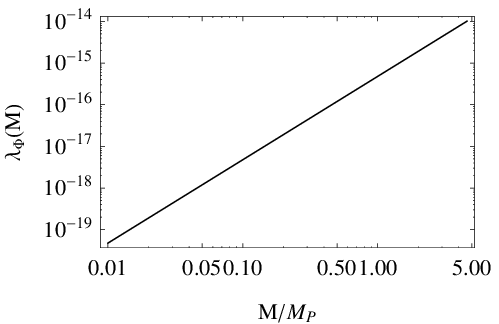}       
\end{center}
\caption{
Left and right panels show the U(1)$_X$ gauge coupling ($g_X(M)$) and 
 the inflaton quartic coupling ($\lambda_\Phi(M)$) as a function of $M/M_{P}$, respectively, 
 for a fixed $x_H=400$. 
}
\label{fig1}
\end{figure}

Let us now present the numerical results of the inflection-point U(1)$_X$ Higgs inflation scenario. 
For the rest of the paper, we employ the e-folding number $N=60$.  
We set $\tilde{g}(v_X)=0$ and choose $\lambda_{\rm mix}(M)$ to be ${\lambda_{\rm mix}}^2 \ll 48 {g_X}^4$, 
   and hence all our results presented in the rest of this section are 
   controlled by only three parameters, $M$, $x_H$ and $v_X$.

\begin{figure}[h]
\begin{center}
\includegraphics[scale =1.53]{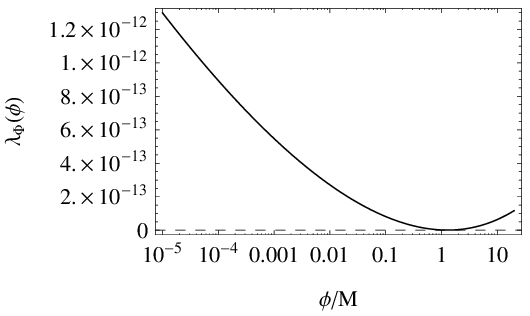} \;\;\;\;
\includegraphics[scale=1.53]{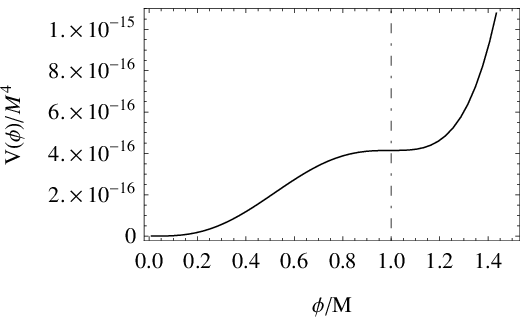}       
\end{center}
\caption{
Left panel shows the RG running of the U(1)$_X$Higgs/inflaton quartic coupling 
   plotted against the normalized energy scale $\phi/M$.
Here we have fixed $M = M_P$ and $x_H=400$, which corresponds to 
   $g_X(M)=8.756 \times 10^{-4}$, $Y_M(M)=2.478 \times 10^{-3}$, 
   and $\lambda_\Phi(M) \simeq 4.770 \times10^{-16}$.  
The dashed horizontal line  indicates $\lambda=0$. 
Right panel shows the corresponding RG improved inflaton potential, 
	where the inflection-point-like point appears at $\phi=M$. 
}
\label{fig2}
\end{figure}

In Fig.~\ref{fig1}, we show the U(1)$_X$ gauge coupling (left) and the inflaton quartic coupling (right)  
   at $\phi=M$ as a function of $M$ (see Eqs.~(\ref{FEq1}) and (\ref{FEq2})).  
Here, we have fixed $x_H=400$, which is motivated by the LHC phenomenology 
  to be discussed in the next section.

In Fig.~\ref{fig2}, we plot the running quartic coupling $\lambda_\Phi(\phi)$ (left) and 
   the RG improved effective inflaton/Higgs potential (right).  
Here we have fixed $M=M_P$ and $x_H=400$, 
   which corresponds to $g_X(M)=8.760 \times 10^{-4}$, $Y_M(M)=2.103 \times 10^{-3}$, 
   and $\lambda_\Phi(M) \simeq 4.770 \times10^{-16}$. 
In the left panel, the dashed line indicates $\lambda_\Phi=0$. 	
In the right panel, we see the inflection-point-like behavior of the inflaton potential 
   around $\phi=M$, marked with a dashed-dotted vertical line.

Here let us look at the inflationary predictions of our scenario. 
The prediction for the tensor-to-scalar ratio ($r$) is given by Eq.~(\ref{FEqR}). 
For the upper bound on $M < 5.67 M_P$, the resultant tensor-to-scalar ratio is given by 
  $r < 3.047 \times 10^{-5} $, which is too small to test in the future experiments.    
However, as discussed in the previous section, the inflection-point inflation predicts 
   the running of the spectral index to be $\alpha \simeq - 2.742 \times 10^{-3}$, 
   independently of $M$. 
This predicted value is within the reach of future precision measurements \cite{RunningSpectral}.

We now consider the particle mass spectrum of the model at low energies.  
As we have discussed, the condition of the almost vanishing $\beta_{\lambda_\Phi} (M)$ 
   leads to the relation of $Y_M(M) \simeq 32^{1/4} g_X(M)$. 
The low energy mass spectrum of the $Z^\prime$ boson and the right-handed neutrinos 
   are obtained by extrapolating the couplings to low energies. 
From Eq.~(\ref{FEq2}) with the upper bound on $M< 5.67 M_P$, we can see $g_X(M) \ll 1$ 
  and hence the RG running effects of the gauge and Yukawa couplings are negligible 
  and $m_{N}/m_{Z'}\simeq 0.84$ remains almost the same at low energies. 
On the other hand, the RG evolution of the inflaton quartic coupling significantly changes 
  its value at low energies (as shown in the left panel of Fig.~\ref{fig2}), 
  since its beta function is controlled by the gauge and Yukawa couplings. 
Let us approximately solve the RG equations in  Eq.~(\ref{ApproxRGE}).  
Since $g_X(M)$, $Y_M(M) \ll 1$, the solutions to their RG equations are approximately given by 
\bea
g_X(\mu)   &\simeq& g_X(M) +\beta_{g}(M) \ln \left[\frac{\mu}{M}\right]  ,\nonumber \\ 
Y_M(\mu)  &\simeq& Y_M(M)+ \beta_{Y}(M) \ln \left[\frac{\mu}{M}\right], 
\label{gYatVEV}
\eea
where $\beta_{g}(M)$ and  $\beta_{Y}(M)$ are their beta-functions evaluated at $\mu=M$.  
Hence, the beta-function of the quartic coupling is approximately described as 
\bea
\beta_{\lambda_\Phi}(\mu) &\simeq& 96 g^4(\mu) - 3 Y^4(\mu)  \nonumber \\ 
&\simeq &  4 \left( 96g_X(M)^3 \beta_{g}(M)-3Y_M(M)^3\beta_{Y}(M)\right) \ln \left[\frac{\mu}{M}\right] 
\nonumber \\ 
&\simeq& M\beta^\prime_{\lambda_\Phi}(M) \ln \left[\frac{\mu}{M}\right] 
 \simeq 16 \lambda_\Phi(M) \ln \left[\frac{\mu}{M}\right],  
\label{betaapprox}
\eea 
where we have used $M\beta^\prime_{\lambda_\Phi}(M) \simeq 16 \lambda_\Phi(M)$ in Eq.~(\ref{Cond1}). 
Then we obtain the approximate solution to the RG equation as 
\bea
\lambda_\Phi(v_X)  &\simeq& 8 \lambda_\Phi(M)\left(\ln \left[\frac{v_X}{M}\right] \right)^2  
\nonumber\\
 &\simeq& 3.868 \times 10^{-15} \Big(\frac{M}{M_{P}}\Big)^2 \left(\ln \left[\frac{v_X}{M}\right] \right)^2,
\label{lambdaatVEV}
\eea 
where we have used Eq.~(\ref{FEq1}) and $v_X \ll M$. 
Using $m_{Z^\prime} = 2 g(v_X) v_X \simeq 2 g(M) v_X$, 
   the mass ratio of the U(1)$_X$ Higgs/inflaton to the $Z^{\prime}$ boson is given by 
\bea
\frac{m_\phi}{m_{Z'}} \simeq 2.911 \times 10^{-6}  \left(\frac{M}{M_P} \right)^{2/3} 
  {\left(87+256x_H+164 x_H^2\right)^{1/6}} \ln \left[\frac{M}{v_X}\right].
\label{ratiophiz} 
\eea

\section{LHC Run-2 Constraints}
If kinematically allowed, the $Z^\prime$ boson in the minimal U(1)$_X$ model can be produced at the LHC. 
The ATLAS and the CMS collaborations have been searching for a narrow resonance 
   with di-lepton final states at the LHC Run-2 and set the upper limits of 
   a $Z^\prime$ boson production cross section~\cite{ATLAS:2016, CMS:2016}.  
In the analysis by the ATLAS and the CMS collaborations, 
   the so-called sequential SM $Z^\prime$ ($Z^\prime_{SSM}$) model \cite{ZpSSM} 
   has been considered as a reference model, where the $Z^\prime$ boson 
   has the same couplings with the SM fermions as the SM $Z$ boson.    
In this section, we interpret the current LHC constraints on the sequential $Z^\prime$ boson  
   into the U(1)$_X$ $Z^\prime$ boson to identify an allowed parameter region. 
Then, we examine a consistency of the inflection-point inflation scenario 
   with the LHC Run-2 constraints.

We first calculate the cross section for the process $pp \to Z^\prime +X \to \ell^{+} \ell^{-} +X$. 
The differential cross section with respect to the invariant mass $M_{\ell \ell}$ of the final state di-lepton 
   is given by
\begin{eqnarray}
 \frac{d \sigma}{d M_{\ell \ell}}
 =  \sum_{q, {\bar q}}
 \int^1_ \frac{M_{\ell \ell}^2}{E_{\rm CM}^2} dx
 \frac{2 M_{\ell \ell}}{x E_{\rm CM}^2}  
 f_q(x, Q^2) f_{\bar q} \left( \frac{M_{\ell \ell}^2}{x E_{\rm CM}^2}, Q^2
 \right)  {\hat \sigma} (q \bar{q} \to Z^\prime \to  \ell^+ \ell^-) ,
\label{CrossLHC}
\end{eqnarray}
where $f_q$ is the parton distribution function for a parton (quark) ``$q$'', 
  and $E_{\rm CM} =13$ TeV is the center-of-mass energy of the LHC Run-2.
In our numerical analysis, we employ CTEQ6L~\cite{CTEQ} for the parton distribution functions 
   with the factorization scale $Q= m_{Z^\prime}$. 
Here, the cross section for the colliding partons is given by 
\bea 
{\hat \sigma}(q \bar{q} \to Z^\prime \to  \ell^+ \ell^-) =
\frac{\pi}{1296} \alpha_X^2 
\frac{M_{\ell \ell}^2}{(M_{\ell \ell}^2-m_{Z^\prime}^2)^2 + m_{Z^\prime}^2 \Gamma_{Z^\prime}^2} 
F_{q \ell}(x_H),  
\label{CrossLHC2}
\eea
where $\alpha_X=g_X^2/(4 \pi)$,  the function $F_{q \ell}(x_H)$ are 
\bea
   F_{u \ell}(x_H) &=&  (8 + 20 x_H + 17 x_H^2)  (8 + 12 x_H + 5 x_H^2),   \nonumber \\
   F_{d \ell}(x_H) &=&  (8 - 4 x_H + 5 x_H^2) (8 + 12 x_H + 5 x_H^2) 
\label{Fql}
\eea
for $q$ being the up-type ($u$) and down-type ($d$) quarks, respectively, 
   and the total decay width of $Z^\prime$ boson is given by 
\bea
\Gamma_{Z'} = 
 \frac{\alpha_X}{6} m_{Z^\prime} 
 \left[ F(x_H) + 3 \left( 1- \frac{4 m_N^2}{m_{Z^\prime}^2} \right)^{\frac{3}{2}} 
 \theta \left( \frac{m_{Z^\prime}}{m_N} - 2 \right)  \right] 
\label{width}
\eea
with $ F(x_H)=13+ 16 x_H  + 10 x_H^2 $. 
By integrating the differential cross section over a range of $M_{\ell \ell}$ set by the ATLAS and the CMS analysis, 
  respectively, we obtain the cross section to be compared with the upper bounds 
  obtained by the ATLAS and the CMS collaborations.

In interpreting the 2016 results by the ATLAS and the CMS collaborations 
   into the U(1)$_X$ $Z^\prime$ boson case, 
   we follow the strategy in \cite{OO2} (see also \cite{OO1} for the minimal $B-L$ model). 
In the paper, the cross section for the process $pp \to Z^\prime_{SSM} +X \to \ell^{+} \ell^{-} +X$ 
   is calculated and the resultant cross sections are scaled by a $k$-factor 
   to match with the theoretical predictions presented in the ATLAS and the CMS papers. 
With the $k$-factor determined in this way, the cross section 
   for the process $pp \to Z^\prime+X \to \ell^{+} \ell^{-} +X$  
   is calculated to identify an allowed region for the model parameters 
   of $\alpha_X$, $x_H$ and $m_{Z^\prime}$.    
See Ref.~\cite{OO2} for the details of the strategy and the $k$-factors. 
Our analysis in this section is exactly the same as that in this reference.

\begin{figure}[h]
\begin{center}
\includegraphics[scale =1.3]{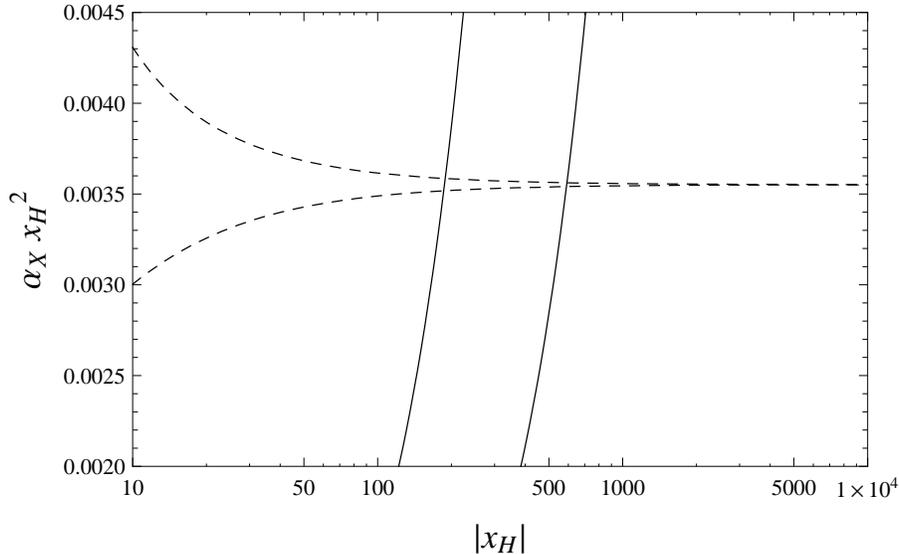} 
\end{center}
\caption{
The upper bounds on $\alpha_X \; x_H^2$ as a function of $x_H$ from the CMS results 
  on search for a narrow resonance from the combined di-electron and di-muon channels \cite{CMS:2016}.   
The lower and upper dashed lines correspond to $x_H >0$ and $x_H < 0$, respectively. 
Here we have fixed $m_{Z^\prime}=4$ TeV. 
The diagonal solid lines depict Eq.~(\ref{FEq1}) for $M=M_P$ (left) and $M=0.01 M_P$ (right), 
  along which the successful inflection-point inflation is achieved.  
} 
\label{fig3}
\end{figure}

For $m_{Z^\prime}=4$ TeV, we show in Fig.~\ref{fig3} the upper bounds 
  on $\alpha_X \; x_H^2$ as a function of $x_H$ from the CMS results 
  on search for a narrow resonance from the combined di-electron and di-muon channels \cite{CMS:2016}. 
The lower and upper dashed lines correspond to $x_H >0$ and $x_H < 0$, respectively.     
The upper bounds from the ATLAS results \cite{ATLAS:2016} are found to be very similar to 
   but slightly weaker than those from the CMS results, and we have shown only the CMS results.  
As we can see from the cross section formula, the dashed lines approach each other for a large $|x_H|$.    
The diagonal solid lines depict Eq.~(\ref{FEq1}) for $M=M_P$ (left) and $M=0.01 M_P$ (right), 
  along which the successful inflection-point inflation is achieved.  
For the diagonal solid lines with a fixed $M$, the results for $x_H >0$ and $x_H < 0$ are indistinguishable.   
From this figure, we find an upper bound on $x_H \lesssim 200$ and $x_H \lesssim 600$, respectively,  
   for $M=M_P$ (left) and $M=0.01 M_P$ (right).  
Note that even though the successful inflection-point inflation requires the U(1)$_X$ gauge coupling 
   to be very small, this scenario can still be tested at the LHC when $|x_H| \gg 1$, 
   in other words, the U(1)$_X$ gauge symmetry is oriented towards the SM hyper-charge direction.

\section{Constraints from the Big Bang Nucleosynthesis}
Let us now consider a reheating scenario after inflation to connect our inflation scenario 
  with the Standard Big Bang Cosmology. 
This occurs via inflaton decay into the SM particles during the inflaton oscillates around its potential minimum.   
We estimate the reheating temperature ($T_R$) as
\begin{eqnarray}
    T_R \simeq 0.55 \left(\frac{100}{g_*}\right)^{1/4} \sqrt{\Gamma_\phi M_P} , 
\label{TR}
\end{eqnarray} 
where $\Gamma_\phi$ is the inflaton decay width into the SM particles. 
For the successful Big Bang Nucleosynthesis (BBN), we impose a model-independent  lower bound on 
   the reheating temperature as $T_R \gtrsim 1$ MeV.

In the Higgs potential of Eq.~(\ref{Higgs_Potential}),  
    a mass matrix between the inflaton ($\phi$) and the SM Higgs boson ($h$)  
    is generated after the U(1)$_X$ symmetry and the electroweak symmetry breaking: 
\begin{eqnarray}
{\cal L}  \supset -
\frac{1}{2}\begin{bmatrix}h  & \phi\end{bmatrix}
\begin{bmatrix} 
m_h^2 &  \lambda_{\rm mix} v_h v_X \\ 
 \lambda_{\rm mix} v_h v_X & m_{\phi}^2
\end{bmatrix} 
\begin{bmatrix} h \\ \phi \end{bmatrix}, 
\label{massmatrix}
\end{eqnarray} 
where $m_{\phi} = \sqrt{2 \lambda_\Phi} v_X$, and $m_h = \sqrt{2 \lambda_{H}} v_{h} = 125$ GeV.  
We diagonalize the mass matrix by 
\begin{eqnarray}
\begin{bmatrix} h \\ \phi \end{bmatrix}   =
\begin{bmatrix} \cos\theta &   \sin\theta \\ -\sin\theta & \cos\theta  \end{bmatrix} \begin{bmatrix} \phi_1 \\ \phi_2 
\end{bmatrix}  ,
\end{eqnarray} 
where $\phi_1$  and $\phi_2$ are the mass eigenstates. 
The relations among the mass parameters and the mixing angle ($\theta$) are the following: 
\bea
&2 v_h v_X  \lambda_{\rm mix} = ( m_h^2 -m_\phi^2) \tan2\theta,   \nonumber  \\
 &m_{\phi_1}^2 = m_h^2     - \left(m_\phi^2  - m_h^2 \right) \frac{\sin^2\theta}{1-2 \sin^2\theta} , \nonumber \\
 &m_{\phi_2}^2 = m_\phi^2 + \left(m_\phi^2 - m_h^2 \right) \frac{\sin^2\theta}{1-2 \sin^2\theta} \ .
\label{mixings} 
\eea
Since the inflaton is much lighter than the $Z^\prime$ boson and the heavy neutrinos, it decays to the SM particles mainly through the mixing with the SM Higgs boson. 
We calculate the inflaton decay width as 
\bea 
   \Gamma_{\phi_2} = \sin^2\theta \times  \Gamma_h(m_{\phi_2}) , 
\label{width_phi}   
\eea
where $\Gamma_h(m_{\phi_2})$ is the SM Higgs boson decay width 
   if the SM Higgs boson mass were $m_{\phi_2}$.

There are constraints on the mixing angle. 
Firstly, we have imposed $\lambda_{\rm mix}^2 \ll {48 g_X^4}$ to neglect the contribution 
  of the $\lambda_{\rm mix}$ to $\beta_\Phi$. 
Another constraint on the mixing angle is from requiring positive definiteness 
  of mass squared eigenvalues of the mass matrix in Eq.~(\ref{massmatrix}), 
  which leads to $\lambda_{\rm mix}^2 < 4 \lambda_{H} \lambda_{\Phi}$. 
We find that the latter constraint is more severe and requires $\theta \ll 1$.  
Hence $\phi_1$ and $\phi_2$ are mostly the SM Higgs and the U(1)$_X$ Higgs mass eigenstates, respectively.

\begin{figure}[h]
\begin{center}
\includegraphics[scale =1.4]{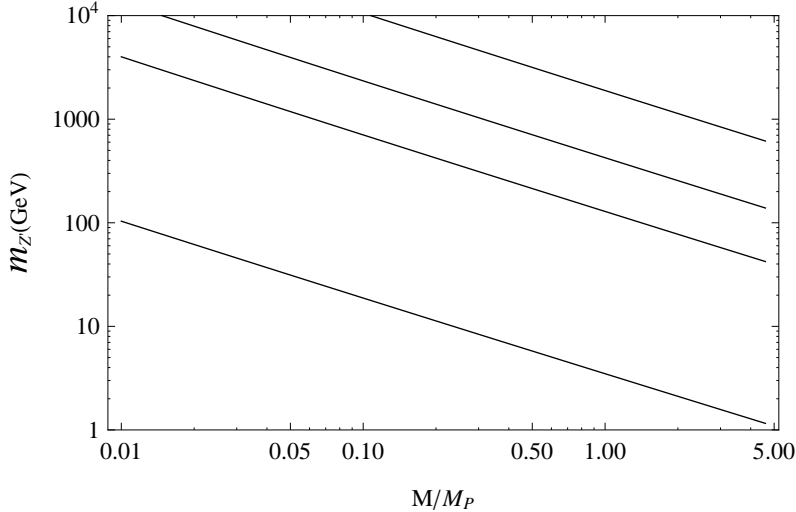} 
\end{center}
\caption{
The contours corresponding to the reheating temperatures $T_R= 1$ MeV, $1$ GeV, $100$ GeV, and $1$ TeV,  
   from bottom to top, for $x_H=400$ and $\xi = 0.1$. 
} 
\label{fig4}
\end{figure}

In the following analysis we parameterize $\lambda_{\rm mix}^2  = 4 \lambda_{H} \lambda_{\phi}  \xi$ 
  with a new parameter $0< \xi <1$. 
From Eq.~(\ref{mixings}), we obtain 
\bea
\theta^2 \simeq  \xi \left(\frac{m_{\phi}}{m_h}\right)^2,
\label{mixingangle}
\eea  
where we have used $m_\phi^2 \ll m_h^2$ from Eq.~(\ref{ratiophiz})
  for the parameter region we are interested in, namely  
  $m_{Z^\prime}={\cal O}$(1 TeV). 
We also find that $m_{\phi2} \simeq m_{\phi}\sqrt{1-\xi}$. 
From Eqs.~(\ref{ratiophiz}), (\ref{TR}), (\ref{width_phi})  and (\ref{mixingangle}), 
   we can express the reheating temperature as a function of $M$,  $m_{Z^\prime}$, $x_H$ and $\xi$. 
To simplify our analysis, let us fix $x_H=400$ and $\xi = 0.1$.  
In Fig.~\ref{fig4}, we show the contours corresponding to $T_R= 1$ MeV, $1$ GeV, $100$ GeV, and $1$ TeV, 
   from bottom to top.

\section{Conclusions}
From a theoretical point of view, if the inflaton value is trans-Planckian, 
   effective operators suppressed by the Planck mass 
   could significantly affect the inflaton potential during inflation, 
   and hence the inflationary predictions. 
To avoid this problem, we may consider the SFI, 
   where the inflaton value during inflation is smaller than the Planck mass. 
In this case, the inflection-point inflation is an interesting possibility to realize a successful slow-roll inflation
   when inflation is driven by a single scalar field. 
To realize the inflection-point-like behavior for the RG improved effective $\lambda \phi^4$ potential, 
   the running quartic coupling $\lambda(\phi)$ must exhibit a minimum 
   with an almost vanishing value in its RG evolution, namely $\lambda(\phi_I) \simeq 0$ and $\beta_{\lambda}(\phi_I) \simeq 0$, where $\beta_{\lambda}$ is the beta-function of the quartic coupling.

From a particle physics perspective, it is more compelling to consider an inflationary scenario, 
   where the inflaton field plays another important role. 
We may consider a general Higgs model, namely the gauge-Higgs-Yuakwa system, 
   and identify the Higgs field as inflaton.  
In this case, the conditions, $\lambda(\phi_I) \simeq 0$ and $\beta_{\lambda}(\phi_I) \simeq 0$, 
   lead to a relation among the gauge, the Yukawa and the Higgs quartic couplings. 
Using the relation and requiring the inflationary predictions to be consistent with the Planck 2015 results, 
   we have found that all the couplings (at $\phi_I$) depend only on $\phi_I$. 
Hence, the low energy mass spectrum of the model is uniquely determined by only two free parameters, 
   $\phi_I$ and the inflaton/Higgs VEV, and 
    the inflationary predictions are complementary to the low energy mass spectrum. 
It is also interesting that the inflection-point inflation provides a unique prediction 
    for the running of the spectral index $\alpha \simeq  - 2.7 \times 10^{-3}$, 
    which can be tested in the future experiments.

We have investigated the inflection-point inflation in the context of the minimal U(1)$_X$ extended SM, 
    where the anomaly-free extra gauge symmetry is defined as a linear combination of 
    the SM hyper-charge and the gauged $B-L$ groups. 
Identifying the U(1)$_X$ Higgs field with the inflaton, 
    we have obtained a prediction for the mass spectrum for the $Z^\prime$ boson, 
   the right-handed neutrinos, and the U(1)$_X$ Higgs boson 
   as a function of $\phi_I$, $x_H$, and the inflaton/Higgs VEV.  
Even though the successful inflection-point inflation requires the U(1)$_X$ gauge coupling to be very small, 
   we have found that the $Z^\prime$ boson with mass of a few TeV  
   can be explored at the LHC Run-2 when the direction of the U(1)$_X$ symmetry 
   is oriented towards the SM hyper-charge, or equivalently $|x_H|\gg1$.  
This is in sharp contrast to the inflection-point inflation scenario 
   in the minimal U(1)$_{B-L}$ extended SM previously investigated in Ref.~\cite{OR_SFI}. 
We have also considered the reheating after inflation and found a large portion of parameter 
   space which can reheat the universe sufficiently high.

\section*{Acknowledgements}
S.O. would like to thank the Department of Physics and Astronomy at the University of Alabama 
  for hospitality during her visit for the completion of this work. 
She would also like to thank FUSUMA Alumni Association at Yamagata University 
  for travel supports for her visit to the University of Alabama. 
This work is supported in part by the United States Department of Energy (Award No.~DE-SC0013680).

\appendix
\section{RG equations in the minimal U(1)$_X$ model} 
\label{Sec:U(1)'_RGEs}
In this Appendix we list the one-loop RG equations for the couplings which are used in our analysis.  
See Appendix in Ref.~\cite{OOT1} for a complete list.
The RG equations for the gauge couplings at the one-loop level are given by  
\begin{eqnarray}
\mu \frac{d g_1}{d\mu} &=& \frac{g_1}{(4\pi)^2} 
	\left[ 12\left(\frac{1}{6}g_1+x_q\tilde{g} \right)^{\!\!2} + 6\left(\frac{2}{3}g_1+x_u\tilde{g} \right)^{\!\!2} 
		+ 6\left(-\frac{1}{3}g_1+x_d\tilde{g} \right)^{\!\!2} 
	\right. \nonumber \\
	&+& \left.
	 4\left(-\frac{1}{2}g_1+x_\ell\tilde{g} \right)^{\!\!2}
		+ 2\left(x_\nu\tilde{g} \right)^2 + 2\left(-g_1+x_e\tilde{g} \right)^2
	     + \frac{2}{3}\left(\frac{1}{2}g_1+x_H\tilde{g} \right)^{\!\!2} 
		+ \frac{1}{3}\left(x_\Phi\tilde{g} \right)^2 \right], 
 \nonumber \\  
\mu \frac{d g_X}{d\mu} &=& \frac{g_X^3}{(4\pi)^2} 
	\left[ 12x_q^2 + 6x_u^2 + 6x_d^2 + 4x_\ell^2+ 2x_\nu^2 + 2x_e^2
		 + \frac{2}{3}x_H^2 + \frac{1}{3}x_\Phi^2 \right], 
    \nonumber \\  
\mu \frac{d \tilde{g}}{d\mu} &=& \frac{1}{(4\pi)^2} 	\left[ \tilde{g} \left\{ 
		12\left(\frac{1}{6}g_1+x_q\tilde{g} \right)^{\!\!2} + 6\left(\frac{2}{3}g_1+x_u\tilde{g} \right)^{\!\!2} 
		+ 6\left(-\frac{1}{3}g_1+x_d\tilde{g} \right)^{\!\!2} 
	\right. \right. \nonumber \\
	&+&  4\left(-\frac{1}{2}g_1+x_\ell\tilde{g} \right)^{\!\!2}
		+ 2\big(x_\nu\tilde{g} \big)^2 + 2\big( - g_1 + x_e\tilde{g} \big)^2
	\left.
		 + \frac{2}{3}\left(\frac{1}{2}g_1+x_H\tilde{g} \right)^{\!\!2} 
		+ \frac{1}{3}\left(x_\Phi\tilde{g} \right)^2 \right\} \nonumber \\
	&+& 2g_X^2 \!
	\left\{ 12x_q \! \left(\frac{1}{6}g_1+x_q\tilde{g} \right) + 6x_u \!\left(\frac{2}{3}g_1+x_u\tilde{g} \right) 
		+ 6x_d \!\left(-\frac{1}{3}g_1+x_d\tilde{g} \right) 
	\right. \nonumber \\
	&+& 4x_\ell\left(-\frac{1}{2}g_1+x_\ell\tilde{g} \right)
		+ 2x_\nu \big(x_\nu \tilde{g} \big) + 2x_e\big(-g_1+x_e\tilde{g} \big)
	\nonumber \\
	&+&\left. \left.
		 \frac{2}{3}x_H \left(\frac{1}{2}g_1+x_H\tilde{g} \right) 
		+ \frac{1}{3}x_\Phi \left(x_\Phi\tilde{g} \right) \right\} \right]. 
\label{Eq:RGE_gauge}  
\end{eqnarray}
Here, $x_f$ is a U(1)$_X$ charge of a corresponding fermion ($f$) in Table~\ref{table1}.  
For example, $x_q=(1/6) x_H + (1/3) x_\Phi$, and $x_e=-x_H-x_\Phi$. 
For the RGEs for the Majorana Yukawa couplings at the one-loop level we have 
\bea
\mu \frac{d Y_M^i}{d\mu} = \frac{Y_M^i}{(4\pi)^2} 
	\Bigg[ (Y_M^i)^2 + \frac{1}{2} \sum_{j=1}^3 (Y_M^j)^2 
			+ \left( 12x_\nu^2 - 6x_\Phi^2 \right) \! \left( \tilde{g}^2 + g_X^2 \right) \Bigg]. 
\label{Eq:RGE_Yukawa}  
\end{eqnarray} 
Finally, the RGEs for the scalar quartic couplings are given by 
\begin{eqnarray}
\mu \frac{d \lambda_\Phi}{d\mu} &=& \frac{1}{(4\pi)^2} 
	\Bigg[ \lambda_\Phi \Big\{ 20\lambda_\Phi + 2\sum_{i=1}^3 (Y_M^i)^2 
							- 12\big( x_\Phi \tilde{g} \big)^2 - 12\big( x_\Phi g_X \big)^2 \Big\} 
	\Bigg. \nonumber \\
	&+&\Bigg.
		 2\lambda_{\rm mix}^2 - 4\sum_{i=1}^3 (Y_M^i)^4 
		+ 6 \Big\{ \big( x_\Phi \tilde{g} \big)^2 + \big( x_\Phi g_X \big)^2 \Big\}^2 
	\Bigg], 
	 \nonumber \\  
\mu \frac{d \lambda_{\rm mix}}{d\mu} &=& \frac{1}{(4\pi)^2} 
	\left[ \lambda_{\rm mix} \Bigg\{ 12\lambda_H + 8\lambda_\Phi + 4\lambda_{\rm mix} 
							+ 6y_t^2 + \sum_{i=1}^3 (Y_M^i)^2 \Bigg. 
	\right. \nonumber \\
	&-& \left.
		 \frac{9}{2}g_2^2 - 6\left( \frac{1}{2}g_1+x_H\tilde{g} \right)^{\!\!2} - 6\big(x_\Phi\tilde{g} \big)^2 
		- 6\big( x_H g_X \big)^2 - 6\big( x_\Phi g_X \big)^2 
	 \right\} \nonumber \\
	&+& \left.
		12 \left\{ \left( \frac{1}{2}g_1 + x_H\tilde{g} \right) \!\! \big( x_\Phi \tilde{g} \big) 
						+ \big( x_H g_X \big) \! \big( x_\Phi g_X \big) \right\}^{2} 
	\right].
\label{Eq:RGE_lambda}
\end{eqnarray}



\end{document}